\documentclass[12pt,a4paper,twoside]{article}
\pdfoutput=1
\usepackage{amsmath}
\usepackage{graphicx}
\usepackage[usenames,dvipsnames]{color}
\usepackage[colorlinks=true, urlcolor=blue, linkcolor=blue, 
citecolor=blue]{hyperref}

\usepackage{bbm}                                                  
\usepackage{cprotect}
\usepackage{slashed}

\usepackage{cite}

\usepackage[top=0.8in,bottom=1in,left=0.9in,right=0.9in]{geometry}
\def\acal{{\cal A}}
\def\ccal{{\cal C}}
\def\fcal{{\cal F}}
\def\gcal{{\cal G}}
\def\lcal{{\cal L}}
\def\mcal{{\cal M}}
\def\ncal{{\cal N}}
\def\ocal{{\cal O}}
\def\scal{{\cal S}}

\def\half{\frac12}
\def\re{{\bf Re}}
\def\su#1{{SU(#1)}}
\def\ui{U(1)}
\def\up#1{^{\left( #1 \right) }}
\def\mati{{\mathbbm1}}
\def\inv#1{\frac1{#1}}
\def\vevof#1{\left\langle #1 \right\rangle}
\def\then{{\quad\Rightarrow\quad}}
\def\mev{\hbox{MeV}}
\def\gev{\hbox{GeV}}
\def\tev{\hbox{TeV}}
\def\fm{\fcal}

\def\phitd{\tilde\phi^\dagger}
\def\gdm{\gcal_{\rm DM}}
\def\pmns{V}
\def\msc{m_\Phi}
\def\mfe{m_\Psi}
\def\cw{c_{W}}
\def\sw{s_{W}}
\def\lx{\lambda_x}
\def\gsim{\mathrel{\rlap{\lower4pt\hbox{\hskip1pt$\sim$}}
    \raise1pt\hbox{$>$}}}                
\newcommand{\bea}{\begin{eqnarray}}  \newcommand{\eea}{\end{eqnarray}}
\newcommand{\beq}{\begin{equation}}  \newcommand{\eeq}{\end{equation}}
\newcommand{\bit}{\begin{itemize}}   \newcommand{\eit}{\end{itemize}}

\begin{document}
\thispagestyle{empty}

\hfill{\sc UG-FT-317/15}

\vspace*{-2mm}
\hfill{\sc CAFPE-187/15}

\vspace{32pt}

\begin{center}

\textbf{\Large 
A realistic model for Dark Matter interactions \\[1ex] 
in the neutrino portal paradigm
}

\vspace{50pt}
Vannia Gonz\'alez Mac{\'\i}as,$^a$
Jos\'e I. Illana,$^b$
Jos\'e Wudka$^a$
\vspace{16pt}

\textit{$^a$Department of Physics \& Astronomy}\\ 
\textit{ University of California, Riverside, CA 92521, USA}\\
\vspace{10pt}
\textit{$^b$CAFPE and Departamento de F{\'\i}sica Te\'orica y del Cosmos}\\
\textit{Universidad de Granada, E-18071 Granada, Spain}\\
\vspace{16pt}

\texttt{
vannia.gonzalez-macias@ucr.edu,
jillana@ugr.es, 
jose.wudka@ucr.edu}

\end{center}

\vspace{30pt}

\begin{abstract}

We discuss a simple extension of the Standard Model (SM)  that provides an explicit realization of the dark-matter (DM) neutrino-portal paradigm. The dark sector is composed of a scalar $ \Phi $ and a Dirac fermion $ \Psi $, with the latter assumed to be lighter than the former. These particles interact with the SM through the exchange of a set of heavy Dirac fermion mediators that are neutral under all local SM symmetries, and also under the dark-sector symmetry that stabilizes the $ \Psi $ against decay. We show that this model can accommodate all experimental and observational constraints provided the DM mass is below $\sim 35\, \gev $ or is in a resonant region of the Higgs or $Z$ boson. We also show that if the dark scalar and dark fermion are almost degenerate in mass, heavier DM fermions are not excluded. We note that in this scenario DM annihilation in the cores of astrophysical objects and the galactic halo produces a monochromatic neutrino beam of energy $ \mfe $, which provides a clear signature for this paradigm. Other experimental signatures are also discussed.

\end{abstract}

\newpage

\section{Introduction}

Dark matter (DM) presents one of the most interesting aspects of physics beyond the Standard Model (SM).  The most compelling DM paradigm assumes that it consists of one or more particles with very weak couplings to the SM \cite{Goodman:1984dc,Feng:2010gw}, and having the correct abundance to explain the CMB observations \cite{Ade:2013zuv}. This hypothesis has been probed extensively using direct \cite{Akerib:2013tjd,Aprile:2012nq,Ahmed:2009zw} and indirect detection \cite{Choi:2015ara,Aartsen:2014hva,Ackermann:2015zua,Aharonian:2006wh,Aguilar:2013qda,Adriani:2008zr} experiments, and in collider processes \cite{Khachatryan:2014rra,ATLAS:2012ky,Goodman:2010ku,Bai:2010hh}. To date, no evidence of DM effects in any of these experiments has been confirmed.\footnote{ Some direct-detection experiments \cite{Bernabei:2013xsa} have published evidence of DM effects, but these are inconsistent with other experimental results and remain controversial.}

The above paradigm has given rise to a large number of publications proposing specific models of DM-SM interactions \cite{DM-SM-models,Jungman:1995df,Bertone:2004pz}, as well as phenomenological descriptions of these interactions based on the effective Lagrangian approach (see, for example, \cite{Goodman:2010ku,Belanger:2008sj,Duch:2014xda,Crivellin:2014qxa,Crivellin:2014gpa}). In particular, reference \cite{Macias:2015cna} describes a possible scenario that ensures  naturally small direct and indirect detection signals, without compromising the relic abundance inferred from CMB experiments. This scenario is based on the assumption that interactions between the dark and SM sectors are mediated by one or more Dirac fermions $\fm$, assumed neutral under all dark and SM symmetries, except fermion number. In addition, the dark sector is assumed to contain (at least) one fermion $ \Psi $ and one scalar  $ \Phi $ that have the same (non trivial) transformations under a symmetry group $\gdm $, whose nature we will not need to specify; we only assume that all SM particles are singlets under $ \gdm $, which ensures that the lightest dark-sector particle will be stable and so serve as a DM candidate. 

The general considerations in \cite{Macias:2015cna}, however, do not necessarily demonstrate the full phenomenological viability of this scenario since there might be additional effects of the $\fm$ that are not associated with the dark sector, and which may set further constraints. In this paper we will construct the simplest specific model that realizes such a scenario and investigate the implications of existing and projected experimental restrictions on the model parameters. We show that despite the high precision constraints available, there are significant regions of parameter space that are still allowed. We will also show that the model has a distinctive identifying feature: the presence of a monochromatic neutrino signal generated by DM annihilation in astrophysical objects.

\section{Effective theory considerations}

The main motivation for the scenario described above comes from the observation that a dark sector that contains scalars $ \Phi $ and fermions $ \Psi $ allows for the presence of an effective interaction with the SM of the form
\beq
\ocal\up5 = (\bar\Psi \Phi)(\phitd \ell) 
\label{eq:d5}
\eeq
where $ \ell $ and $ \phi $ denote, respectively, the isodoublets for a left-handed SM lepton and SM scalar ($\tilde\phi=i\sigma^2\phi^*$); this dimension-5 operator can be generated at tree-level by the exchange of a fermion $ \fm $. To understand the implications of the coupling (\ref{eq:d5}) note that in the unitary gauge $ \ocal\up5 = (\bar \Psi \nu_L \Phi) (v+H)/\sqrt{2} $, where $\nu_L$ and $H$ denote  a SM left-handed neutrino and the Higgs field, respectively, and $ v \simeq 246\,\gev $ is the electroweak scale. Within the $\fcal$-mediated paradigm this operator describes the strongest interactions between the SM and the dark sector, which always involve a neutrino: this is a neutrino portal scenario (neutrino portals have been studied in related contexts for example in \cite{NeuPortal}). The presence of a $\Psi$-$\Phi$-$\nu$ coupling also implies that the heaviest of the dark particles will promptly decay into the lightest, so there will be a single DM relic despite having a dark sector with two (or more) particles. 

In addition to the above $\fm$-induced coupling, the presence of dark scalars allows for the usual Higgs portal coupling $ |\Phi|^2 |\phi|^2 $. If the dark fermion is heavier than the dark scalar, $ \mfe > \msc $, then $\Phi$ constitutes the DM relic and the physics of the model is dominated by the effects of the Higgs portal coupling, which has been extensively studied in the literature \cite{Silveira:1985rk,McDonald:1993ex,Burgess:2000yq,Bento:2000ah,Holz:2001cb,LopezHonorez:2006gr}. In contrast, if $ \mfe < \msc $, the Higgs portal coupling is secondary to (\ref{eq:d5}) and the phenomenology is completely different; for example, the leading interactions relevant for direct detection are produced by the dimension  6 effective operators
\beq
 |\phi|^2 \bar\Psi \Psi  \qquad (\phi^\dagger\!\!\stackrel\leftrightarrow{D_\mu}\!\!\phi ) \left(\bar\Psi \gamma^\mu P_{L,R}\Psi \right)
 \qquad ( \bar\ell \gamma_\mu \ell ) (\bar\Psi \gamma^\mu P_{L,R}\Psi )
 \label{eq:d6}
\eeq
that are generated at one loop\footnote{Current-current operators involving quarks or right-handed leptons are only generated at $\ge2$ loops.} by the $ \fm$. This will be discussed in full detail below.

The fact that the $ \fm $ create interactions (\ref{eq:d5}) at tree-level and (\ref{eq:d6}) at one loop is what  allows for the required relic abundance to be obtained within the constraints of direct and indirect detection experiments, without fine-tuning. However, there will be additional restrictions: the $ \fm $ can also mix with the neutrinos and so generate deviations within the SM of processes such as $W$ and $Z$ decays involving the $ \nu_L $, as well as certain $W$-mediated processes such as meson decay. As noted previously, these constraints must be studied in the context of a specific model, since only then we can obtain the relations between the effective interactions such as (\ref{eq:d5},\ref{eq:d6}) with  those involving the vector bosons, such as $  \overline{\nu_L} \slashed Z  \nu_L $,  and so determine the viability of the scenario.

\section{The model}

In this section we describe the details of the specific model we consider.  We take the simplest case where the dark sector contains one scalar $\Phi$ and one fermion field $\Psi$, transforming under a global symmetry under which all SM fields are singlets.\footnote{This represents the simplest possibility realizing our scenario; the model can be easily generalized to include additional scalars and fermions, and possible gauge symmetries.} We will assume that $ \msc > \mfe $, so that the fermion is stable.  In addition we assume that the only interaction between the dark fields and the SM is through the exchange of three Dirac fermions $\fm$, neutral under the dark and the SM symmetries. Finally, we require that lepton number be conserved (except for possible instanton effects).\footnote{The effects of a small Majorana mass for the $\fm$ will be briefly discussed below.}

The Lagrangian of the model is:
\bea
\lcal &=& \bar\ell i \slashed{D} \ell + \overline{ e_R } i \slashed D e_R + \bar\Psi(i \slashed\partial - \mfe) \Psi + \bar\fm(i \slashed\partial - M) \fm + |\partial\Phi|^2 - \msc^2 |\Phi|^2 \cr
&& \quad - \left( \bar\ell Y\up e e_R \phi + \bar\ell Y\up\nu \fm \tilde\phi + \bar\Psi  z^\dagger \fm \Phi +  {\rm H.c.} \right)-\lambda_x|\Phi|^2|\phi|^2
\label{eq:lag}
\eea
where $ \ell_i$ and $e_{R\,i}$ denote, respectively, the left-handed SM lepton isodoublets and right-handed isosinglets ($i=1,2,3$ is a family index); $ \phi$ is the SM scalar isodoublet; $ \Psi$ and $ \Phi$ are the dark fermion and scalar fields; and $ \fm_i$ are the (Dirac) neutral fermion mediators.  $M$ is the $ 3 \times 3 $ Hermitian mass matrix for the $\fm$, while the  Yukawa couplings $Y\up e, Y\up\nu$ are general $3\times 3$ complex matrices.

The theory has three conserved quantities associated with the lepton sector: lepton number $L$, `dark' number $D$, and (global) hypercharge $Y$. The corresponding charges are:
\bea
\begin{array}{l|cccccc}
  & \ell & e_R & \phi & \fcal & \Psi & \Phi \cr
\hline
L & 1 & 1 & 0 & 1 & 1 & 0 \cr
D & 0 & 0 & 0 & 0 & 1 & 1 \cr
Y & -\half &-1 & \half & 0 & 0 &0 
\end{array}
\eea

The Lagrangian (\ref{eq:lag}) contains 54 parameters, but we also have the freedom to make $U(3)$ (family) rotations for the  $ e_R,\, \ell,\, \fcal$ and $\ui$ for $\phi,\, \Phi,\, \Psi$, with 3 of these transformations corresponding to the $L$, $D$ and $Y$ transformations. There remain 27 physical parameters \cite{Santamaria:1993ah} that we choose as follows. We use the $U(3)$ rotation freedom for $ \fm $ to diagonalize $M$ and make $z$ real. Then we use the $U(3)$ rotation freedom for $\ell$ and $e_R$ so that $m_e = (v/\sqrt{2})Y\up e$, the charged-lepton mass matrix, is diagonal and positive. There remains a $ \ui^3 $ freedom  that we use to eliminate 3 phases in $ Y\up\nu $, that we use together with the standard polar decomposition to write
\beq
 Y\up\nu =\frac{\sqrt{2}}v \pmns \eta U M\,,
\eeq
where $\pmns$ will correspond to the Pontecorvo-Maki-Nagakawa-Sakata (PMNS) matrix, $\eta $ is a positive diagonal matrix, and $ U \in \su3 $.

Finally, we replace $ \fm$ and $ \nu $ by the fields $N$ and $n_L$ which diagonalize the mass matrix:
\bea
\fcal &=& U^\dagger (\ccal U_L N_L -  \scal n_L + U_R N_R ) \cr
\nu &=& \pmns(\scal U_L N_L + \ccal n_L)
\label{eq:fieldrot}
\eea
where
\beq
\ccal = \inv{\sqrt{\mati + \eta^2 }} \qquad \scal = \frac\eta{\sqrt{\mati + \eta^2 }}
\label{eq:ccal}
\eeq
with the unitary matrices $U_{L,R} $ are chosen such that
\beq
U_R^\dagger  U M U^\dagger \ccal^{-1} U_L = M_N = {\rm diagonal}\,.
\label{eq:Mdiag2}
\eeq
In this basis the $n_L$ are massless left-handed fermions that correspond to the SM neutrinos, and $(N_L,N_R)$ form a set of Dirac fermions with mass matrix $M_N$. The interaction terms then become:
\bea
-\lcal_{\rm int} &=&   \left(H/v \right) \bar e m_e e + \lx |\phi|^2 |\Phi|^2 + \frac g{2\cw} \bar e \slashed Z ( 2 \sw^2 - P_L) e \cr
&& \qquad + \left[\bar\Psi  z^T U^\dagger( \ccal U_L N_L - \scal n_L + U_R N_R)  \Phi +  {\rm H.c.} \right] \cr
&& \qquad  + \frac g{\sqrt{2}} \left[ \bar e \slashed W \pmns  (\ccal n_L + \scal U_L N_L) + {\rm H.c.} \right] \cr
&& \qquad + \left(H/v \right) \left[  \bar N_R M_N U_L^\dagger \scal (\ccal n_L + \scal U_L N_L )   +  {\rm H.c.} \right] \cr
&& \qquad   + \frac g{2\cw}  (\bar n_L \ccal +  \bar N_L U_L^\dagger \scal) \slashed Z (\ccal n_L + \scal U_L N_L)\,.
\eea
The number of physical parameters is 3 for $ m_e,\,M,\,z$ and $ \eta$ each, plus 1 for $\mfe,\,\msc,\,\lx $ each, plus 4 for $\pmns$ and 8 for $U$, for a total of 27.

A mass terms for the $n_L$ can be generated by adding to (\ref{eq:lag}) a small Majorana mass term for the $ \fcal $:
\beq
\lcal \to \lcal - \half \left( \fcal^T {\sf C} \mcal \fcal + {\rm H.c.} \right) 
\eeq
where ${\sf C}$ is the usual charge-conjugation matrix. Assuming $\mcal$ is small we see immediately from (\ref{eq:fieldrot}) that this term generates a Majorana mass for the $n$:\footnote{The $N$ also acquire a small Majorana mass, that we will assume small compared to $M_N$.}
\bea
\lcal_{n-\rm Maj} 
&=& - \half n_L^T {\sf C}  \left(  \scal^T U^{*} \mcal  U^\dagger \scal\right) n_L + {\rm H.c.}  \,, 
\label{eq:maj}
\eea
which has the typical inverse-seesaw \cite{Wyler:1982dd,Mohapatra:1986bd,Ma:1987zm} form. The matrix $ \mcal $ can then be adjusted to meet the experimental constraints on the neutrino masses and mixing angles. We will not further investigate this aspect of the model since it is not relevant for the discussion of the DM phenomenology, which is the main objective of this paper.

\subsection{Non-standard interactions}

The presence of the $ \fcal $ modifies the couplings of the charged and light neutral leptons to the SM bosons, as well as couplings involving the heavy neutral leptons $N$:
\bit
\item $Z$ couplings:
\beq
  - \frac g{2\cw} \left[ \bar n_L \ccal^2  \slashed Z n_L 
 + \bar N_L U_L^\dagger \scal^2 U_L \slashed Z N_L 
+ \left(\bar n_L \ccal \scal U_L\slashed Z N_L +{\rm H.c.} \right) \right]\,.
\label{eq:Znsm}
  \eeq
No flavor changing neutral currents coupled to the $Z$ boson are induced since $ \ccal $ is diagonal  (cf. Eq. \ref{eq:ccal}); however, the couplings are no longer universal since the elements of $\ccal$ are not necessarily all equal.  Note also from the definition that each element of $\ccal$ is positive and smaller than one.\footnote{
If a light neutrino mass matrix is introduced, e.g. as in (\ref{eq:maj}), and diagonalized, the $nnZ$ couplings will not, in general, remain diagonal.
}
\item $W$ couplings:
 \beq
  - \frac g{\sqrt{2}} \left[ \bar e \slashed W \pmns  \ccal n_L  + \bar e \slashed W \pmns \scal U_L N_L+ {\rm H.c.} \right] \,.
\label{eq:Wnsm}
\eeq
Note that in addition to the usual unitary mixings generated by the PMNS matrix $\pmns$, these couplings also involve the non-universal  $ \ccal $ factors.

\item Yukawa couplings:
\beq
- \left(H/v \right) \left[  \bar N_R M_N U_L^\dagger \scal \ccal n_L    +  \bar N_R M_N U_L^\dagger \scal^2 U_L N_L   +  {\rm H.c.} \right]\,.
\label{eq:Ynsm}
\eeq
If  $m_H$ is greater than the mass of any of the $N$, there is a contribution to the Higgs invisible decay width generated by the first term.

\item DM-SM interactions (neutrino portal):

The coupling of the DM fields with the $n$ and $N$ take the form
\beq
 \bar\Psi  z^T U^\dagger  \scal n_L   \Phi -  \bar\Psi  z^T U^\dagger( \ccal U_L N_L  + U_R N_R)  \Phi +  {\rm H.c.}
\label{eq:dmsm}
\eeq
As emphasized above, the first term containing $n_L$ represents the leading couplings between the dark and SM sectors, which justifies our denoting this a ``neutrino portal'' scenario. The presence of the $ \Psi$-$\Phi$-$n_L$ coupling implies that whenever $ \msc > \mfe $ the scalar field will decay promptly into the fermion and a neutrino:\footnote{If $ \mfe > \msc$ it is the fermions that decay.} this model, while having a multi-component dark sector, has a single component DM relic. However, the presence of the $ \Phi $ is essential for generating the leading DM-SM interactions.

\eit

\subsection{\texorpdfstring{$Z$-DM}{ZDM} couplings}

\begin{figure}[th]
\centerline{
\includegraphics{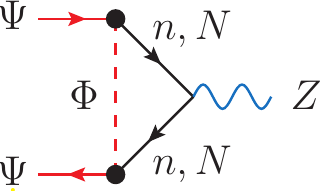}
}
\caption{Interaction of the $Z$ boson and the DM fields induced at one loop.\label{fig:PPZ1L}}
\end{figure}

The DM interactions with the $Z$ boson are induced at one loop (fig. \ref{fig:PPZ1L}), yet they represent some of the leading couplings in direct detection processes. In addition, though in general they generate  small corrections in the annihilation cross section, they may produce important resonant effects when $m_\Psi\simeq m_Z/2$.  

A straightforward evaluation of the diagrams in figure~\ref{fig:PPZ1L} gives (at zero external momenta)
\beq
-i \frac g{2 \cw} \left( \epsilon_L \gamma^\mu P_L+ \epsilon_R \gamma^\mu P_R \right)
\label{eq:Zdm}
\eeq
where
\bea
\epsilon_R &=&- \inv{32\pi^2} \sum_{i,j} (z^T M U^\dagger U_R)_i (M_N U_L^\dagger \scal^2 U_L M_N)_{ij} (U_R^\dagger U M z)_j \, f_0(\msc^2, M_{N\, i}^2, M_{N\, j}^2) \cr
\epsilon_L &=&  + \inv{16\pi^2} \sum_{i,j} (z^T U^\dagger U_R)_i (M_N U_L^\dagger \scal^2 U_L M_N)_{ij} (U_R^\dagger U z)_j \, f_1(\msc^2, M_{N\, i}^2, M_{N\, j}^2)\,,
\eea
the $M_{N\, i}$ are the diagonal entries in $M_N$, and
\beq
f_n(a,b,c) = \frac{a^n \ln a}{(a-b)(a-c)} + \frac{b^n \ln b}{(b-a)(b-c)}+ \frac{c^n \ln c}{(c-b)(c-a)}\,.
\label{eq:fun}
\eeq
The couplings (\ref{eq:Zdm}) correspond to the dimension-6 operator $(\phi^\dagger\!\!\stackrel\leftrightarrow{D_\mu}\!\!\phi) (\bar\Psi \gamma^\mu P_{L,R}\Psi) $ in (\ref{eq:d6}); as argued on general grounds in \cite{Macias:2015cna}, they are loop-generated within the neutrino portal scenario.

\subsection{\label{sec:D-DM} \texorpdfstring{$H$-DM}{HDM} couplings}

\begin{figure}[th]
\centerline{
\includegraphics{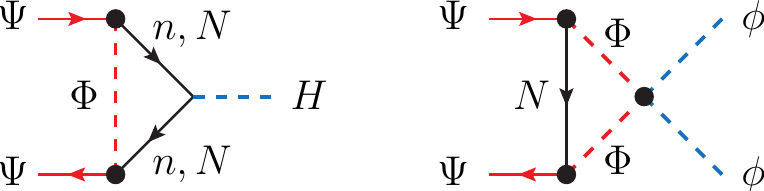}
}
\caption{Interactions of the Higgs boson and the DM fields induced at one loop.\label{fig:PPH1L}}
\end{figure}

As in the case of the $Z$ boson, the $H$-DM interactions are induced at one loop (fig.~\ref{fig:PPH1L}), and correspond to the effective operator $ \bar\Psi \Psi |\phi|^2$ in  (\ref{eq:d6}), which is also loop-generated. Note the presence of a contribution involving the Higgs portal coupling $ \lambda_x $. A tedious calculation gives
\bea
&& - \frac i{8\pi^2 v}   \sum_{i,j} \Biggl\{ 
\left[ (z^T U^\dagger U_R)_i (M_N U_L^\dagger \ccal  U z)_j P_R + (z^T U^\dagger\ccal U_L M_N)_i (U_R^\dagger U z)_j P_L \right] \cr && ~ \times
\left[ (M_N U_L^\dagger \scal^2 U_L M_N )_{ij} f_1(\msc^2 , M_{N\, i}^2,  M_{N\, j}^2 )   - \half \lambda_x v^2 \delta_{ij} 
f_1(\msc^2, \msc^2 ,  M_{N\, i}^2) \right] \Biggr\} \cr &&
\label{eq:Hdm}
\eea
where $f_1$ is defined in (\ref{eq:fun}). These couplings are relevant in direct detection processes and for the annihilation cross section in the resonant region $ \mfe \simeq m_H/2 $. 

\section{Quasi-degenerate heavy fermions}
\label{sec:DHF}

In the rest of the paper we will concentrate on the special case where $ |\eta|\ll1 $ and the mass states $N$ are almost degenerate:  $M \simeq \Lambda [\mati+\ocal (\eta^2)] $. In this case  (\ref{eq:Mdiag2}) implies that, up to $\ocal (\eta^2)$ corrections, $ U_{L} = U_{R}=U $, $\ccal \simeq \mati - \eta^2/2 $, $\scal \simeq \eta $ and $ M \simeq \Lambda(\mati-\frac{1}{2}U^\dagger\eta^2 U)$.

Then the $Z$-DM couplings simplify considerably:
\beq
\epsilon_R \simeq \frac{|\eta U z|^2 }{16\pi^2} \ln \frac\Lambda\msc \,, \qquad
\epsilon_L \simeq \frac{|\eta U z|^2}{16\pi^2} 
\label{eq:erlh}
\eeq
while the Higgs-DM couplings (\ref{eq:Hdm}) reduces to
\beq
 - \frac{i\Lambda}{8\pi^2 v}   \left\{ |\eta U z|^2   + \lambda_x |z|^2 \frac{v^2}{\Lambda^2 }  \ln \frac\Lambda\msc  \right\}\,.
\label{eq:Hdmh}
\eeq
Also, the neutrino interactions with the dark sector (\ref{eq:dmsm}) reduce to
\beq
\Phi^\dagger \bar n_L (\eta U z) \Psi + {\rm H.c.}
\label{eq:dmsm-degen}
\eeq
As a consequence, the observables of interest (the cross sections relevant for relic abundance and indirect detection calculations) will  depend on $\lx,\, \eta,\,U$ and $z$ only through the two real combinations $|\eta U z|$ and $\lx |z|^2$. 

\subsection{Electroweak constraints}

The tightest restrictions on the model parameters are derived from the decays of the $W$ and $Z$ gauge bosons, and from the limits on the invisible decay of the Higgs boson. The constraints below are presented for the case of quasi-degenerate heavy fermions.

\paragraph{$Z$ invisible decay}
\label{sec:Z-inv}

The addition of singlet Dirac fermions $N$ to the SM generate non-universal, though flavor diagonal, neutrino couplings to the $Z$ proportional to $\ccal^2 $ (cf. eq.~\ref{eq:Znsm}). In particular, the invisible $Z$ width will be proportional to tr$(\ccal^4) $.  The experimental result $\Gamma(Z\to {\rm inv})=499.0 \pm 1.5\, \mev$ \cite{Agashe:2014kda} for the invisible width of the $Z$ then generates the most stringent bound on the parameters of the model:
\beq
\Gamma(Z\to {\rm inv}) = \inv3{\rm tr}(\ccal^4) \Gamma_{\rm SM}(Z\to {\rm inv}) \then \sum_i  \eta_i^2 < 0.014 \quad (3\sigma)
\label{eq:Z_limits}
\eeq
where the heavy $N$ are taken degenerate and $ \eta_i $ are the diagonal elements of the diagonal matrix $ \eta $. Additional constraints can be derived if $Z$ decays involving the $N$ singlets are kinematically allowed (a region of parameter space that we will not consider since it is disfavored by the relic abundance requirements).

\paragraph{$W$-mediated decays}

\begin{figure}[th]
\centerline{\includegraphics{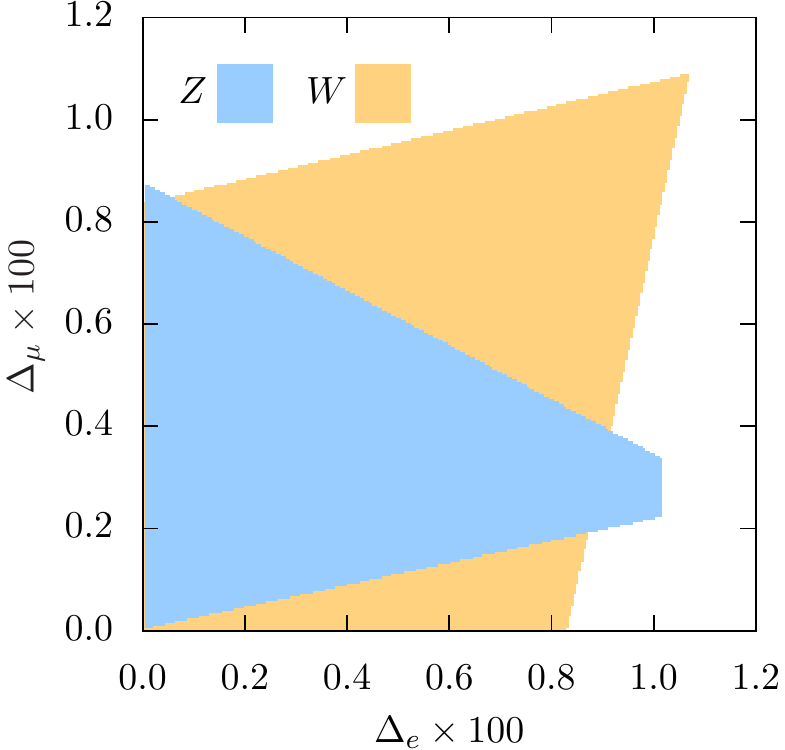}}
\caption{Limits on the parameters $ \Delta_{e,\,\mu}$ (cf. eq. \ref{eq:Deltas}) obtained from  (\ref{eq:Z_limits}) and (\ref{eq:W_limits}), respectively labeled $Z$ and $W$.
\label{fig:Deltas}}
\end{figure}

The charged current interactions within the leptons and the $W$ boson are also modified (cf. eq. \ref{eq:Wnsm}), so that the vertex involving a charged lepton $e_{L\,i}$ and a neutrino mass eigenstate $n_{L\,j}$ contains a factor $(V\ccal)_{ij}$, where $i,j$ are flavor indices. The width of a $W$-mediated decay such as 
$\tau\to e\nu\bar{\nu}$, $\tau\to\mu\nu\bar{\nu}$ and $\pi\to \mu\nu$ will be proportional to $1 - \Delta_i$ where
\beq
\Delta_i = 1-\left( V \ccal^2V^\dagger \right)_{ii} \simeq \left(V \eta^2 V^\dagger \right)_{ii}
= \sum_j |V_{ij}|^2\eta_j^2>0
\label{eq:Deltas}
\eeq
for  $i=e,\,\mu,\,\tau$ (no sum over $i$). Using the current uncertainties we then find (at $3\sigma$)
\bea
\tau \to \mu\nu\bar\nu:&& | 0.8223\,\Delta_\mu - 0.1958\,\Delta_e | \le 0.0069\cr
\tau \to e\nu\bar\nu : && | 0.1777\,\Delta_\mu - 0.8042\,\Delta_e | \le 0.0067\cr
\pi\to \mu\nu: && |\Delta_\mu - \Delta_e| \le 0.010\,,
\label{eq:W_limits}
\eea
leading to the constraints $|\Delta_{e,\,\mu}| \le 0.011$, that are weaker than those from (\ref{eq:Z_limits}), as figure~\ref{fig:Deltas} shows. Other constraints can be derived from experiments testing neutrino mixings \cite{Antusch:2014woa,Parke:2015goa}, which also lead to weaker bounds on the $\eta$ parameters than those in (\ref{eq:Z_limits}).

\paragraph{Electroweak precision data}
A global fit to the electroweak precision data sets model independent limits on lepton mixings of new to SM fermions \cite{deBlas:2013gla}. In particular, one obtains limits on individual $\eta_i$ that are stronger than those in (\ref{eq:Z_limits}) for the first two families, but the third one is of the same order. However, since  the restrictions to our DM model  depend on $|\eta U z|$ that involves an arbitrary unitary rotation, these limits will not impose further constraints.

\paragraph{Higgs invisible decay}

In the following we will be interested in the case where the $N$ are heavier than the Higgs. In this case $H$ may still decay to the DM relics, $H\to\Psi\bar\Psi$, when $\msc<m_H/2$,  via the loop-suppressed couplings described in section \ref{sec:D-DM}. In the quasi-degenerate heavy fermions approximation (\ref{eq:Hdmh}) the partial decay width is given by
\beq
\Gamma(H\to \Psi\bar\Psi)=\frac{v^2 m_H}{512 \pi^5 \Lambda^2}\left( 1-\frac{4 \mfe^2}{m^2_H}\right)^{3/2}
\left[ |\eta U z|^2\frac{\Lambda^2}{v^2}  + \lx |z|^2 \ln \frac\Lambda\msc \right]^2\,.
\label{eq:Hdec}
\eeq
Latest results from the ATLAS experiment at the LHC \cite{Aad:2015pla,Aad:2015txa,CMS:2015naa} report an upper bound $ \Gamma(H \to {\rm inv}) < 2.2\, \mev $ at a $90\%$ C.L. so that, for $ m_H \gg \mfe $,
\beq
\frac v\Lambda
\left| |\eta U z|^2\frac{\Lambda^2}{v^2}  + \lx |z|^2 \ln\frac\Lambda\msc\right|< 1.7\,.
\label{eq:H_limits}
\eeq

\subsection{Numerical calculations}
\label{sec:num_calc}
In its full generality the model (\ref{eq:lag}) contains 20 undetermined parameters (excluding the charged lepton masses $m_e$, the PMNS matrix $\pmns$, and not including Majorana masses). In performing numerical calculations a full scan of this parameter space is time consuming, but also unnecessary if we are not interested in  quantities involving the heavy $N$ and we adopt the quasi-degenerate scenario described above. In this case the relevant parameters are the masses $ \mfe,\, \msc$, the heavy mass scale  $ \Lambda $ and the coupling combinations $|\eta U z|$ and $\lx |z|^2$, as noted above. In practice we have chosen $\sim2.4\times 10^8$ points in the 5-dimensional parameter space $\{\mfe,\, \msc,\, \Lambda,\, |\eta U z|,\,\lx |z|^2\}$ within the ranges
\bea
&
1\,\gev	\leq \mfe			\leq 80\, \gev \,, \quad
1.01 \,\gev\leq \msc \leq  320 \,\gev\,, \quad
200\, \gev	\leq \Lambda		\leq 10\, \tev \,,  & \cr 
&&\cr
& |\eta U z| \leq 0.24\,,\quad -1.2 \le \lx |z|^2 \leq 1.2\,. &
\label{eq:ranges}
\eea
We analyzed two possibilities for the mass spectrum in the dark sector,  a quasi-degenerate spectrum $m_\Phi<m_\Psi+10\,\gev$ and a non-degenerate spectrum $m_\Phi\ge\mfe+10\,\gev$, with $ \msc > \mfe $ in either case, as required in the scenario we consider. We also exclude data points incompatible with (\ref{eq:Z_limits}) and (\ref{eq:H_limits}). In implementing these constraints we require $|z|\le2$, which is slightly more conservative than the limits $|z_i| < \sqrt{4\pi}$, derived from tree-level unitarity\footnote{We impose tree-level unitarity given our requirement that the model remains perturbative.} for each component of $z$, using the process $ \Psi \fm \to \Psi \fm $. It is worth noting that values of $ |\eta U z|$ above $0.24 $ are excluded by the various constraints discussed above.

\section{Relic abundance}

\begin{figure}[th]
\centerline{
\includegraphics{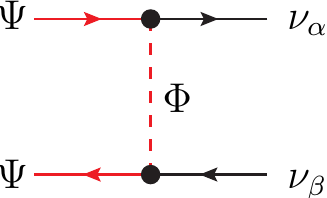}\qquad
\includegraphics{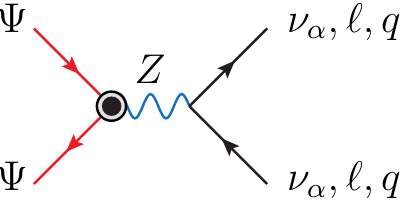}\qquad
\includegraphics{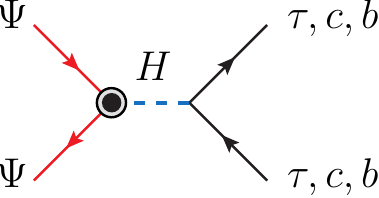}
}
\caption{Leading DM-SM interactions in the annihilation channels. 
\label{fig:PPnn}}
\end{figure}

The leading DM-SM interaction is generated by the (tree-level) exchange of the dark scalars $ \Phi $ (figure \ref{fig:PPnn}) and represent the most important reaction responsible for the equilibration between the dark and standard sectors in the early Universe. This process is produced by the $\Psi$-$\Phi$-$n_L$ interaction (\ref{eq:dmsm}), proportional to $ \eta U z $ to lowest order in $\eta$. It is important to note that even if the dark scalars do not contribute to the relic abundance, their presence is essential for the viability of the model: in the absence of $ \Phi $ all terms in (\ref{eq:dmsm}) would not exist and the SM and dark sectors would decouple.\footnote{For the loop-suppressed terms this follows from a straightforward examination of the possible diagrams.} However, if the dark scalar is slightly heavier than the dark fermion (what we will call the quasi-degenerate case), even though it decays promptly to the dark fermion and a neutrino (\ref{eq:dmsm}), coannihilation processes become important when computing the present density of the relic fermions since the temperature in this case at the equilibration process is higher than the difference in their masses. All of these effects are taken into account in the numerical calculations.\footnote{Coannihilation channels for the equilibration process, such as $\Psi\Phi\to We,\,Z\nu,\,H\nu$, and $\Phi\Phi \to HH,WW,ZZ$ become important when kinematically allowed; all are included in the numerical calculations below.}

The remaining relevant interaction is generated by the one-loop induced $\Psi$-$\Psi$-$H$ coupling in (\ref{eq:Hdmh}), and consists of the $s$-channel exchange of the $H$ boson (figure~\ref{fig:PPnn}). It produces small corrections excepting the resonant region when  $ \mfe \simeq m_H/2 $. A similar interaction generated by the $s$-channel exchange of the $Z$ boson is small but observable in the resonance region $ \mfe \simeq m_Z/2 $.

The leading cross section for $\Psi\bar\Psi\to\nu\bar\nu$ is generated by the left diagram in figure~\ref{fig:PPnn}, and can be calculated using standard techniques; we include the analytic expression in appendix~\ref{sec:xs}. For cold relics ($ T < \mfe $) the leading term in the corresponding thermal average takes the general form $\vevof{\sigma{\rm v}}\simeq \sigma_0(T/\mfe)^n$, where $ \vevof {\rm v} \sim T^{1/2}$ \cite{Kolb:1990vq}. For the present case the leading contribution has $n=0$ (S-wave annihilation) with
\bea
\vevof{\sigma {\rm v}}_{\Psi \bar\Psi \to \nu  \bar\nu }  \simeq (\sigma_0)_{\Psi \bar\Psi \to \nu  \bar\nu } 
&=&  \frac{(v/\Lambda_{\rm eff} )^4}{256 \pi m_\Psi^2} \,, \quad \Lambda_{\rm eff} = 
\frac{v/\sqrt{2}}{|\eta U z|}  \sqrt{\frac{\msc^2+\mfe^2}{\mfe^2}}
\label{eq:snunu}
\eea
where we summed over all final neutrino states.

The $s$-channel exchange of the Higgs boson and $Z$ boson in the $\Psi\bar\Psi\to f\bar f$ annihilation process to heavier fermions (right diagrams in figure~\ref{fig:PPnn}) are small, except in the  resonant cases when $ \mfe \sim m_H/2$ and $\mfe\sim m_Z/2$. The lowest leading dependence in velocity is of second order for the Higgs boson exchange, corresponding to a P-wave annihilation:
\bea
\vevof{\sigma {\rm v}}_{\Psi \bar\Psi \to f\bar f} &=&\frac T\mfe(\sigma_0)_{\Psi \Psi \to f\bar f} \,, \\
(\sigma_0)_{\Psi \bar\Psi \to f\bar f} &=& \frac{N_f \beta_f^3  }{256\pi^5  \Lambda^2} \frac{m_f^2}{\mfe^2} \left[ |\eta U z|^2\frac{\Lambda^2}{ v^2} + \lx |z|^2 \ln\frac{\Lambda}{m_\Phi}\right]^2|\tilde{A}_H|^2 
\label{eq:sff}
\eea
where $N_f$ is the color multiplicity and
\beq
\tilde{A}_H=\frac{m_\Psi^2}{4m_\Psi^2-m_H^2+im_H\Gamma_H} ,\quad \beta_f=\left( 1 - \frac{m_f^2}{\mfe^2} \right)^{1/2}\,.
\label{eq:defs1}
\eeq
While for the $Z$ boson, the lowest leading dependence in velocity is of first order, corresponding to a S-wave annihilation:
\beq
\vevof{\sigma {\rm v}}_{\Psi \bar\Psi \to f\bar f}\simeq\frac{N_f \beta_f  |\eta U z|^4}{512 \pi \mfe^2}|\tilde{A}_Z|^2 \left\{
 \left(1+\ln\frac{\Lambda}{\msc} \right)^2\left[
4g^2_{V_f}+1+\frac{m_f^2}{2m_\Psi^2}(4g_{V_f}^2-1)
\right]  - \frac{2m_f^2}{\mfe^2} \ln\frac{\Lambda}{\msc} \right\}
\label{eq:svPPZff}
\eeq
with
\beq
\tilde{A}_Z=\left(\frac{g}{4\pi c_W}\right)^2\frac{m_\Psi^2}{4m_\Psi^2-m_Z^2+im_Z\Gamma_Z}\,,\quad 
g_{V_f} = T^3_f - 2 Q_f \sw^2
\eeq
where  $T^3_f$ is the weak isospin of $f$, $Q_f$ its charge, and $\sw$ the sine of the weak mixing angle.\\

Adopting the standard freeze-out approximation \cite{Kolb:1990vq}, the relic abundance $\Omega_\Psi $ is given by:
\beq
\Omega_\Psi h^2 = \frac{1.07 \times 10^9}\gev \,(n+1) \frac{ x_f^{n+1} }{g_{\star S}\xi} \,; \quad \xi = \frac{ M_{\rm Pl} \sigma_0 }{\sqrt{g_\star}}
\label{eq:Omapprox}
\eeq
where $M_{\rm Pl}$ denotes the Planck mass, $ g_{\star S} ,\, g_\star $ denote, respectively, the relativistic degrees of freedom associated with the entropy and energy density, $n=0,1$ for S-wave or P-wave processes, respectively,  and
\beq
x_f = \frac\mfe{T_f} = \ln \left(0.152(n+1) \mfe \xi \right) -\left (n+\half\right) \ln \left[ \ln \left(0.152 (n+1)\mfe \xi \right) \right]\,,
\label{eq:xf}
\eeq
with $T_f$ the freeze-out temperature.
This expression for $ \Omega_\Psi $ can now be compared to the result inferred from CMB data obtained by the Planck experiment \cite{Ade:2013zuv}:
\beq
\Omega_{\rm Planck} h^2 = 0.1198\pm0.0026 \quad (3\sigma).
\label{eq:planck}
\eeq

\begin{figure}[th]
$$\includegraphics{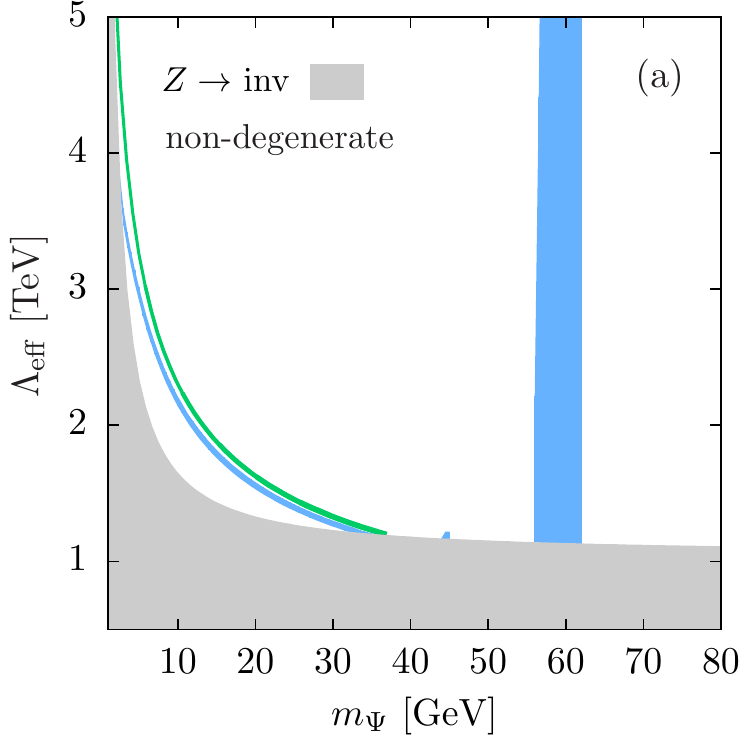}\qquad\includegraphics{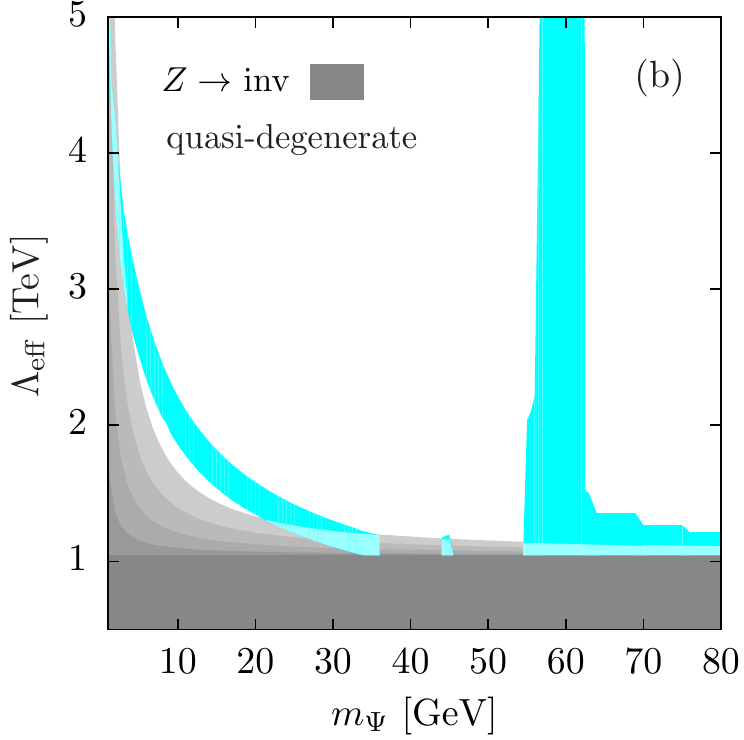} $$
\caption{Constraints on the DM-$\nu$ portal model from the relic abundance $3\sigma$ bounds obtained by the Planck experiment: (a) Allowed regions for the non-degenerate case (blue); for comparison, the green band results from the analytic approximation (\ref{eq:Leff.approx}) valid outside the resonance regions. (b) The cyan areas denote the allowed regions for the quasi-degenerate case; in this case the invisible $Z$ constraint is sensitive to $ \msc$ with the grey bands corresponding  to $\msc =\mfe +\{0,1,3,6,10\}\, \mathrm{GeV}$ (dark to light grey, respectively). The light cyan region corresponds to cases that are allowed for some values of $ \msc $ and not others (see text). Both graphs show the effects of the resonant peaks at $\mfe\sim m_H/2$ and $\mfe\sim m_Z/2$. We set  $|z|=2$ for illustration (see comments at the end of section~\ref{sec:num_calc}); $ \Lambda_{\rm eff} $ is defined in (\ref{eq:snunu}).}
\label{fig:relic_f}
\end{figure}

Outside the resonance region, $ \Omega_\Psi  $ is determined by the $ \Psi \bar\Psi \to \nu\bar\nu $ cross section (\ref{eq:snunu}), so (\ref{eq:planck})  allows only a narrow region in the  $(\mfe, \Lambda_{\rm eff})$ plane (see figure~\ref{fig:relic_f} (a)), which is well approximated by the relation
\beq
\Lambda_{\rm eff} \simeq \sqrt{\frac{m_\Omega }\mfe }\, \tev\,, \quad m_\Omega = 53\, \gev \quad (\hbox{non-resonant~region}).
\label{eq:Leff.approx}
\eeq
with  $\Lambda_{\rm eff}$  defined in (\ref{eq:snunu}).

We have repeated this calculation numerically for the parameter ranges (\ref{eq:ranges}) using the public codes \verb|MicrOmegas| \cite{Belanger:2013oya} and \verb|CALCHEP| \cite{Belyaev:2012qa}. The model implementation for \verb|CALCHEP| was done using the \verb|FeynRules| package \cite{Alloul:2013bka}. We obtain distinctive results for the two cases of the dark mass spectrum, the quasi-degenerate and non-degenerate scenarios. The most stringent constraints are given by the $Z$ invisible width, which projected onto the $\{m_\Psi,\Lambda_{\rm eff}\}$ plane delimit the region 
\beq
\frac{v^2}{2\Lambda_{\rm eff}^2} \left(\frac{\msc^2+\mfe^2}{\mfe^2}\right)<0.058
\label{eq:Z_reg}
\eeq

The results within a 3$\sigma$ range are presented in figure~\ref{fig:relic_f} together with the comparison to the analytic expressions. The allowed regions for the non-degenerate case correspond to the blue areas in figure~\ref{fig:relic_f} (a), the constraints derived from the invisible $Z$ width (grey region) require $ 2.3\le\mfe \le 35\, \gev $ outside the Higgs and $Z$ boson resonance regions, $\mfe\sim m_H/2$ and $\mfe\sim m_Z/2$; the analytic expression corresponds to the upper narrow green band. These results indicate that the neutrino portal scenario favors a light DM model in that case.

Figure~\ref{fig:relic_f} (b) shows the allowed regions (cyan) for the quasi-degenerate case, where  $\mfe < \msc  \leq m_\Psi+10\,\gev$. In this case the constraint derived from the invisible $Z$ width is sensitive to the scalar mass, and correspond to the various grey bands; these only exclude relic masses in the range $35\,\gev\le \mfe\le 53\,\gev$, with heavier relics allowed (in contrast with the non-degenerate case). The lighter cyan region is allowed for certain values of $ \msc $ and excluded for others; for example: if $ \msc \sim \mfe$ (darkest grey band) this whole region is allowed, while if $ \msc \sim \mfe + 10 \, \gev$ (lightest grey band) it is completely excluded. In the quasi-degenerate case the scalars play a more important role in the calculation of the relic abundance because of the importance of the coannihilation processes.

\section{Direct detection}

At present the most stringent limit on spin-independent scattering cross sections of DM-nucleon particles comes from the LUX experiment \cite{Akerib:2013tjd}. In order to derive the corresponding  implications for the model under study we have obtained, using \verb|MicrOmegas|, the DM-nucleon $\Psi \ncal\to \Psi \ncal$ cross sections in the limit where the relative velocity ${\rm v}\to 0$. In this non-relativistic limit the elastic amplitudes are divided into spin-independent interactions, generated by scalar and vector couplings, and spin-dependent interactions, generated by axial-vector couplings. 

The leading interactions between the dark matter and the neutral bosons $Z$ and $H$ are induced at one loop, generated by the diagrams in figures~\ref{fig:PPZ1L} and \ref{fig:PPH1L}, hence, naturally suppressing the interactions of the dark matter with quarks. Both axial-vector and vector and scalar couplings are proportional to $|\eta U z|^2$  or $\lx |z|^2\ln(\Lambda/{m_\Phi})$ in the almost degenerate heavy fermion scenario of section~\ref{sec:DHF}, where only the first parameter combination is affected by the relic abundance constraints (cf.~eq.~\ref{eq:snunu}). Hence both spin-dependent and spin-independent cross sections are roughly of the same order.

The direct-detection spin-independent cross section takes the form 
\beq
\sigma(\Psi \ncal\to\Psi \ncal)_{\rm SI}=\frac4\pi \mu_{\rm red}^2 \left| \frac{Z_{\rm nucl}}{A_{\rm nucl}} \acal_p + \left(1 - \frac{Z_{\rm nucl}}{A_{\rm nucl}} \right) \acal_n \right|^2
\eeq 
where $\acal_{p,n} $ denote, respectively, the amplitudes for proton and neutron scattering in units of inverse mass squared,  $A_{\rm nucl}$ is the atomic number, $Z_{\rm nucl}$ the nuclear charge and $ \mu_{\rm red}$ the $\ncal-\Psi$ reduced mass.

\begin{figure}[th]
\centerline{\includegraphics{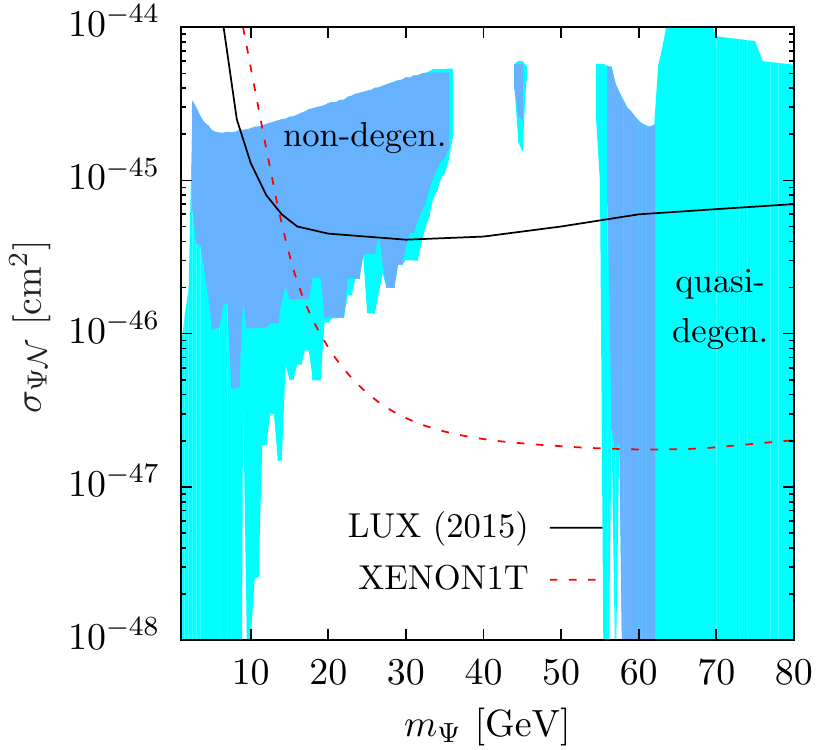}}
\caption{DM-nucleon spin-independent cross sections compatible with the relic abundance and electroweak constraints for the non-degenerate (blue) and quasi-degenerate (cyan) cases. The region above the solid (dashed) lines is (will be) excluded by the LUX (XENON1T) experiments.
\label{fig:DD}}
\end{figure}

Figure~\ref{fig:DD} shows the projection of the numerical results for xenon nuclei to the $(\mfe, \sigma_{\rm SI})$ plane in the quasi-degenerate neutral heavy-fermion scenario, together with the present bounds from LUX  \cite{Akerib:2013tjd} and the expected sensitivity from XENON1T \cite{Aprile:2015zx}. The data points correspond to parameters consistent with the relic abundance and electroweak constraints; blue and cyan refer to the non-degenerate and quasi-degenerate cases of the dark spectrum, respectively. 
The sharp cutoff at $ \mfe \sim 35\, \gev $ is generated by (\ref{eq:Z_limits}). In the non-degenerate case cross sections below $ \sim 10^{-46}\, \rm cm^2$ are excluded for $ \mfe < 35\, \gev $ by (\ref{eq:H_limits}); larger DM masses (and very small cross sections) are allowed only in the quasi-degenerate case. In the vicinity of the Higgs resonance very low cross sections are also allowed; in constrast LUX excludes the $Z$ resonance region, $m_\Psi\sim m_Z/2$. As can be seen from this figure, there are ample regions in parameter space where all restrictions are satisfied.

\section{Indirect detection}

The detection of photons, charged fermions or neutrino final states from the annihilation of DM into SM particles in a dense region of the Universe is known as indirect detection. Within the paradigm studied here neutrinos are the most abundant products; in particular, the annihilation channel gives rise to a monochromatic line in the neutrino energy spectrum from astrophysical sources, which would constitute a ``smoking-gun'' signal for the neutrino-portal scenario under discussion. The expected sources for these neutrino final states are galactic centers, dwarf galaxies, galactic halos, galaxy clusters, and also the cores of the Sun and Earth.

Dark matter particles in the galactic halo have a finite probability to be elastically scattered by a nucleus and become subsequently trapped in the gravitational well of an astronomical object. These DM particles will undergo subsequent scatterings, until they thermalize and concentrate at the core of the object~\cite{Feng:2010gw}. The accumulated DM particles can then annihilate into neutrinos, or other SM particles, that can be detected, among others, in astrophysical high energy neutrino experiments. 
The neutrino spectrum is given by \cite{Gould:1987ww,Lundberg:2004dn,Jungman:1995df}:
\beq
\frac{dN_\nu}{dE_{\nu_i}}\sim\frac{\Gamma_{\Psi\bar\Psi\to \rm SM\,\rm  SM}}{4\pi R^2}\sum_f \mathcal{B}_f \left(\frac{dN_f}{dE}\right)_i
\label{eq:nu-spectrum}
\eeq
where $\Gamma_{\Psi\bar\Psi\to\rm SM\, \rm SM}$ is the DM-DM annihilation rate, $\mathcal{B}_f$ is the branching fraction of channel $f$ and $R$ is the distance from the neutrino source to the detector (Sun-Earth distance or Earth radius for neutrino annihilation in the Sun or Earth, respectively). The function $(dN/dE)_i$ is the differential energy spectrum of neutrino type $i$ at the surface of the object expected from injection of the particles in channel $f$ in their respective cores. Given the small DM velocities, the neutrino spectrum of the $\Psi\bar\Psi\to\nu\bar\nu$ channel in our model is essentially a delta function centered around $E_\nu\simeq \mfe$ that shows up as the above-mentioned monochromatic line in a detector.
 
The annihilation rate is determined by the capture rate $C_\Psi $. When capture and annihilation processes reach equilibrium in a time scale much smaller than the age of the body, one can take $\Gamma_{\Psi\bar\Psi \to \nu \bar\nu} \simeq C_\Psi/2$. The capture rate depends on the elastic scattering cross sections of the DM on nuclei in the Sun or Earth, the DM velocity dispersion and the DM local density \cite{Jungman:1995df}, roughly $C_{\Psi}\propto \sigma_{\Psi\ncal}\times \rho^{DM}_{\rm local}$, where the first factor is generated at one loop by (\ref{eq:Zdm}) and (\ref{eq:Hdm}).

\begin{figure}[th]
\centerline{\includegraphics{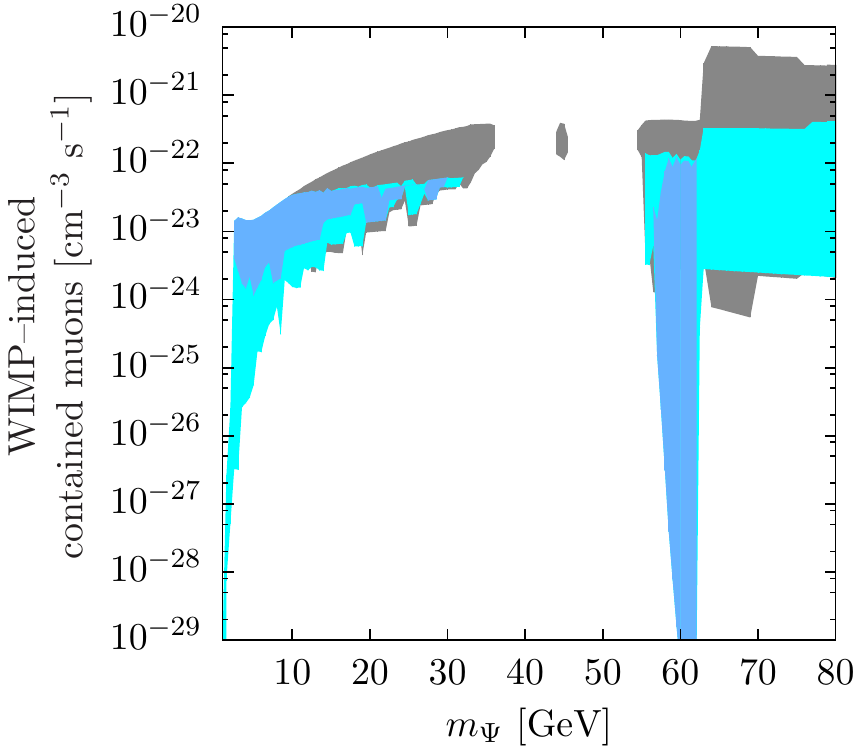}\quad\includegraphics{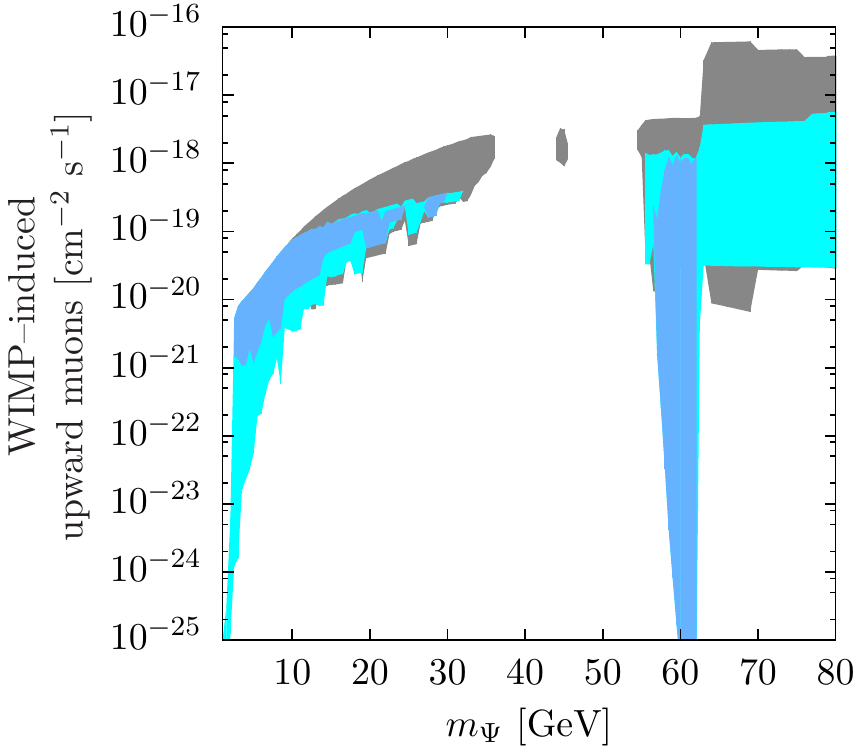}}
\caption{Induced muon rate by neutrinos produced from DM annihilation in the core of the Sun for the non-degenerate (blue) and quasi-degenerate (cyan) cases. The left figure shows the muons produced by neutrinos interacting within the detector (contained), and the right figure shows the induced muon flux by neutrinos interacting with the surrounding material (upward). Grey areas are excluded by LUX.
\label{fig:muonflux}}
\end{figure}

Once produced, neutrinos will travel and interact with the detector; in their passage through the Sun, through space and finally through the Earth, neutrinos will, in general, also interact with matter and/or change its flavor (oscillate). In figure \ref{fig:muonflux} we show the predicted flux of muons produced from the interaction of the neutrinos inside a water Cherenkov detector (contained flux), and the flux of muons produced from the interaction of up-going neutrinos with the rocks surrounding the detector (upward flux), as predicted by our model. We used the \verb|MicrOmegas| package with a neutrino energy threshold of $1\,\rm GeV$. The calculation takes into account not only the dominating process, $\Psi\bar\Psi\to\nu\bar\nu$, but all $\Psi\bar\Psi\to\rm SM\,\rm SM$ channels, and uses the tabulated neutrino spectra functions \cite{Cirelli:2005gh} taking into account effects induced by oscillation and attenuation processes (see for example \cite{Cirelli:2005gh,Blennow:2007tw,Barger:2007hj,Lehnert:2007fv,Erkoca:2009by,Esmaili:2009ks,Esmaili:2010wa,Dasgupta:2012bd}).\footnote{Additional effects such as seasonal variation in the oscillation pattern may alter the total neutrino flux, or the induced-muon flux, when DM annihilates directly into neutrinos, and this may be used to  distinguish between different flavors \cite{Esmaili:2009ks}.  However these effects are small for $\mfe< 100$~GeV and they will be ignored in the following.} Experiments like IceCube or SuperKamiokande use the data collected from the induced-muon flux to derive stronger constraints \cite{Aartsen:2012kia,Choi:2015ara} on the spin-dependent cross sections for DM-proton scattering than those obtained from underground detector experiments, as a result of the hydrogen-rich composition of the Sun. However these limits depend on the DM annihilation final states, which are chosen usually to be the so-called soft ($b\bar b$) or hard channels ($\tau^+\tau^-$) \cite{Choi:2015ara}; also available are limits for the $W^+W^-$ and direct neutrino production channels \cite{Belanger:2015hra}.

The galactic halo, galactic center, galaxy clusters, dwarf galaxy satellites, and other extragalactic unresolved point-sources are also sources of DM annihilation products that may be accessible to indirect detection experiments \cite{Yuksel:2007ac,Bertone:2004pz}. In particular, gamma rays and neutrinos produced as primary or secondary products of DM annihilation will travel essentially undisturbed through space, so the flux of these particles is proportional to the (present time) thermally-averaged, annihilation cross section of  non-relativistic DM relics. 

For our model, neutrinos constitute the dominant flux produced by DM annihilation. The IceCube experiment measures the characteristic anisotropic flux of highly energetic neutrinos for different annihilation channels, including direct annihilation into neutrinos \cite{Aartsen:2014hva}. This experiment is sensitive to neutrino energies above $100\, \gev$ (that  also corresponds to the DM mass) which could impose further restrictions on our model if the dark scalar and dark fermion are quasi-degenerate states, otherwise other constraints require $ 2.3\,\gev<\mfe < 35\, \gev $ or $ \mfe \sim m_H/2 $. In figure \ref{fig:ID} (right) we show the annihilation cross section of $\Psi\bar\Psi$ into neutrinos versus the DM mass $\mfe$ for regions in parameter space that meet the relic abundance constraints (cf. figure {\ref{fig:relic_f}).
There are no significant experimental constraints for neutrino final states in DM annihilation for DM masses below 100~GeV.

Our model also contains subdominant channels of DM annihilation into charged fermions, which arise from the one-loop couplings of the $\Psi$ relics with the $Z$ and $H$ bosons (figures~\ref{fig:PPZ1L} and \ref{fig:PPH1L}). Of these, the $H$-mediated process generates a P-wave annihilation cross section (\ref{eq:sff}), and will be suppressed in the non-relativistic limit (this suppression is much less effective in the relic abundance calculation). In contrast, the $Z$-mediated contribution, while suppressed by small coefficients, generates an S-wave annihilation cross section whose effect on indirect detection processes need not be negligible because of the weaker dependence on the velocity. It also provides the dominant contribution to  $ \Psi \bar\Psi \to b \bar b $. The $H$-mediated and $Z$-mediated cross section are given in (\ref{eq:sff}) and (\ref{eq:svPPZff}) respectively. 

Figure \ref{fig:ID} (left) shows the annihilation cross sections of the process $\Psi\bar\Psi\to b\bar b$ in the non-relativistic limit generated by the $Z$ boson exchange, versus the DM mass $\mfe$, with all points fulfilling the relic abundance and electroweak constraints. The recent Fermi-LAT limit~\cite{Ackermann:2015zua}, obtained by searching for $b\bar b$ annihilation products in several dwarf galaxies having a high ratio of DM to ordinary matter, is also displayed.

\begin{figure}[th]
\centerline{\includegraphics{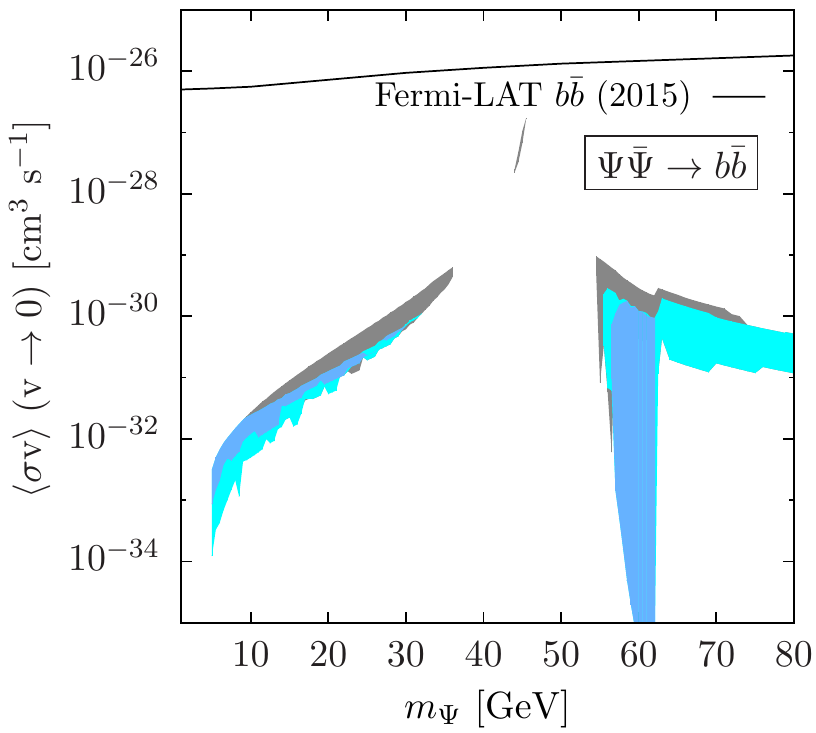}\qquad\includegraphics{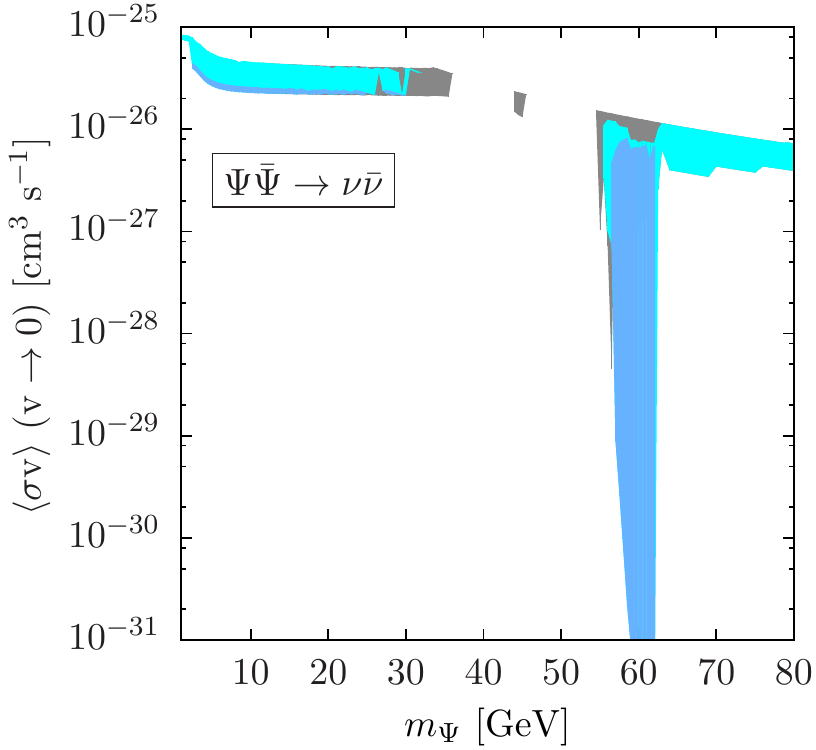}}
\caption{Annihilation cross section into $b$ quarks (left) and $\nu$ (right) final states for the non-degenerate (blue) and quasi-degenerate (cyan) cases. These figures show the allowed region of parameter space from Planck and electroweak constraints at $3\sigma$, together with the latest Fermi-LAT results. Grey areas are excluded by LUX.
\label{fig:ID}}
\end{figure}

\section{Conclusions}

We have studied a specific (minimal) realization of the neutrino-portal scenario where the dark sector contains one scalar $\Phi$ and one fermion $\Psi$, which interact with the SM through the exchange of 3 neutral heavy Dirac fermions $\fm$. In the model, the DM interactions with the SM are loop generated, and so naturally suppressed, except for vertices containing neutrinos which are generated through mixings with the $ \fm$. In particular, DM couplings to all charged SM fermions are small without any fine tuning. 

The relatively large DM-neutrino couplings allow an annihilation cross section large enough to generate the expected relic density, while simultaneously obeying the direct-detection constraints, because of the suppressed couplings to the $Z$ and $H$. The indirect detection constraints are also easily accommodated because in this scenario the main annihilation products are neutrinos, for which the available limits are weak. It is of interest that there are two distinctive scenarios depending on the mass spectrum in the dark sector. If the dark scalars are only sightly heavier than the fermions, coannihilation processes become important in generating the freeze-out of the DM fermions, the lightest and only stable particles, and wide regions of parameter space are allowed. For the case of heavier dark scalars (non-degenerate case) the paramter space is more restricted, favoring a relatively light DM ($\mfe < 35\,\gev$). The main constraints on the model are generated by the $Z$ and $H$ invisible widths. In the case of non-degenerate dark particle states, these constraints restrict the DM mass to lie in the range  $ 2.3\le\mfe \le 35\, \gev $ or near the $H$  resonance region $\mfe\simeq m_H/2$. In contrast, for quasi-degenerate dark scalars and fermions, the electroweak constraints exclude only the relatively narrow range $35\,\gev\le\mfe\le 52\,\gev$, except near the $Z$ resonance. Anyway, the direct detection experiment LUX excludes the $Z$ resonance region. Given that the DM interactions with charged fermions are suppressed, the usual collider signatures (e.g. mono-jet and mono-photon events or missing energy reactions) have  suppressed rates and are beyond reach of the expected experimental sensitivity at the LHC \cite{Abercrombie:2015wmb}.

The cleanest signature of this scenario would be the observation of a monochromatic neutrino line, from both the Sun and the halo, with energy equal to that of the DM mass, but the experimental sensitivity would have to be significantly improved before this can be probed.

\section{Acknowledgments}
We thank F. del \'Aguila and M. P\'erez-Victoria for useful discussions.
This work has been supported in part by a postdoctoral fellowship grant from CONACYT, 
the Spanish MINECO (FPA2013-47836 and Consolider-Ingenio Multidark CSD2009-00064) and Junta de Andaluc{\'\i}a (FQM101, 3048, 6552). J.I.I. acknowledges the University of California, Riverside, for its warm hospitality. 

\appendix
\section{Cross sections}
\label{sec:xs}

In this appendix we provide, for completeness, the expressions of the various cross sections used in the calculation of the relic abundance up to order $\eta^4$.

\paragraph{\bf Neutrino final states}
\begin{align}
\sigma(\Psi \bar\Psi \to \nu_\alpha \bar\nu_\alpha) &=
\frac{|\eta U z |^4}{64 \pi \beta_\Psi s} \bigg\{
\frac{1+2 x(1+x) - \beta_\Psi^2}{(1+x)^2 - \beta_\Psi^2}
+\frac{x}{\beta_\Psi} \ln\left(\frac{1+x - \beta_\Psi}{1+x + \beta_\Psi}\right)
\nonumber\\
&\quad+\frac{1}{2}\re[A_Z(s)] \left[ 
1-x-\left(\frac{2y_\Psi}{x}+\frac{x}{2}\ln\frac{\Lambda}{m_\Phi}\right) \frac{x}{\beta_\Psi}\ln\left( \frac{1+x - \beta_\Psi}{1+x + \beta_\Psi}\right)
\right]
\nonumber\\[1ex]
&\quad+\frac{1}{4}|A_Z(s)|^2 \left[ \frac{1}{4}\left(1+\frac{\beta^2_\Psi}{3}\right)\left(1+\ln^2\frac{\Lambda}{m_\Phi}\right)+2y_\Psi\ln\frac{\Lambda}{m_\Phi}
\right]
\bigg\}
\label{eq:sPPvv}
\\[1ex]
\left. \sigma(\Psi \bar\Psi \to \nu_\alpha \bar\nu_\beta) \right|_{\alpha \not= \beta} &=  \frac{|\eta U z|_\alpha^2|\eta U z|^2_\beta}{64 \pi \beta_\Psi s} \left\{
\frac{1+2 x(1+x) - \beta_\Psi^2}{(1+x)^2 - \beta_\Psi^2}+\frac{x}{\beta_\Psi} \ln \left(\frac{1+x - \beta_\Psi}{1+x + \beta_\Psi}\right)\right\}
\label{eq:sPPvv2}
\end{align}

\paragraph{Charged SM fermion final states}
\beq
\sigma(\Psi\bar\Psi\to f\bar f) = \sigma_H(s) + \sigma_Z(s)
\eeq
with
\begin{align}
\sigma_H(s) &=
\frac{N_f \beta_\Psi\beta_f^3}{1024\pi^5 s}\frac{m_f^2}{v^2}|A_H(s)|^2\left(
\frac{\Lambda}{v}|\eta Uz|^2 +\lx |z|^2\frac{v}{\Lambda} 
\ln\frac{\Lambda}{m_\Phi}
\right)^2
\\
\sigma_Z(s) &=
\frac{N_f \beta_f}{2048\pi \beta_\Psi s}|A_Z(s)|^2\,|\eta U z|^4
\nonumber\\
&\quad\times
\bigg\{
\left(1+\ln^2\frac{\Lambda}{m_\Phi}\right)
\left[
\left(1+\frac{\beta_\Psi^2\beta_f^2}{3}\right)
(4g_{V_f}^2+1)
+4 y_f(1-2y_\Psi)(4g_{V_f}^2-1)
\right]
\nonumber\\
&\qquad
+8y_\Psi\ln\frac{\Lambda}{m_\Phi}
\left[
(1-2y_f)
(4g_{V_f}^2+1)
+4y_f(4g_{V_f}^2-1)
\right]
\bigg\}
\label{eq:sPPff}
\end{align}
In the expressions above we have used
\begin{align}
x&= \frac{2(\msc^2 - \mfe^2)}{s}\,,\quad  y_i = \frac{m_i^2}{s}\,, \quad
\beta_i= \sqrt{ 1 - 4 y_i}\,, 
\nonumber\\
A_Z(s)&=\left(\frac{g}{4\pi c_W}\right)^2\frac{s}{s-m^2_Z+i m_Z\Gamma_Z}\,,\quad
A_H(s)=\frac{s}{s-m^2_H+i m_H\Gamma_H}\ .
\label{eq:defPPff}
\end{align}

\end{document}